\documentclass[a4paper, 12pt]{article}
\usepackage{amsmath}
\usepackage{amssymb}
\usepackage{psfrag}
\usepackage{latexsym}
\usepackage{enumerate}
\usepackage{graphics}
\usepackage{subfigure}
\usepackage{caption}
\input epsf
\usepackage[dvips]{graphicx}
\usepackage{enumerate}
\usepackage[english]{babel}
\usepackage{amsmath,amssymb,amsthm}

\pdfoutput=1

\newcommand{\be}{\begin{equation}}
\newcommand{\ee}{\end{equation}}
\newcommand{\bea}{\begin{eqnarray}}
\newcommand{\eea}{\end{eqnarray}}

\newcommand{\nod}{\noindent}

\newcommand{\ba}{\begin{array}}
\newcommand{\ea}{\end{array}}
\newcommand{\bc}{\begin{center}}
\newcommand{\ec}{\end{center}}

\allowdisplaybreaks

\begin{document}
\title{Mathematical modeling of physical capital using the spatial Solow model}

\author{
\small{Gilberto Gonz\'alez-Parra $^{a,b}$}\footnote{Corresponding author}\\
\small{Email:gcarlos@ula.ve}\\
\small{Benito Chen-Charpentier  $^{b}$}\\
\small{Abraham J. Arenas  $^{c}$}\\
\small{Miguel Diaz-Rodriguez  $^{a}$}\\
\small{$^{\rm a}${Grupo Matem\'atica Multidisciplinar  (GMM), Facultad de Ingenier\'ia.}}\\
\small{Universidad de los Andes, M\'erida, Venezuela.}\\
\small{$^{\rm b}$ {Department of Mathematics, University of Texas at Arlington, Arlington, TX 76019, USA.}}\\
\small{$^{\rm c}$ {Departamento de Matem\'aticas y Estad\'istica, Universidad de C\'ordoba, Monter\'ia, Colombia}}
}

\maketitle \thispagestyle{empty}

\begin{abstract}
This research deals with the mathematical modeling of the physical capital diffusion through the borders of the countries. The physical capital is considered an important variable for the economic growth of a country. Here we use an extension of the economic Solow model to describe how the smuggling affects the economic growth of the countries. In this study we rely on a production function that is non-concave instead of the classical Cobb-Douglas production function. In order to model the physical capital diffusion through the borders of the country, we developed a model based on a parabolic partial differential equation that describes the dynamics of physical capital and boundary conditions of Neumann type. Smuggling is present in many borders between countries and may include fuel, machinery and food. This smuggling through the borders is a problematic issue for the country's economies. The smuggling problem usually is related mainly to a non-official exchange rate that is different than the official rate or subsides. Numerical simulations are obtained using an explicit finite difference scheme that shows how the physical capital diffusion through the border of the countries. The study of physical capital is a paramount issue for the economic growth of many countries for the next years. The results show that the dynamics of the physical capital when boundary conditions of Neumann type are different than zero differ from the classical economic behavior observed in the classical spatial Solow model without physical capital flux through the borders of countries. Finally, it can be concluded that avoiding the smuggling through the frontiers is an important factor that affects the economic growth of the countries.\\
\nod {\bf Keywords: Solow Model; Mathematical modeling; Physical Capital; Numerical Simulation; Neumann boundary conditions.}\\
\nod{PACS 88.05.Lg \and 89.65.Gh}\\
\nod{MSC 35A99 \and 35Q91 \and 37N40}
\end{abstract}

\section{Introduction}
The Solow model presented in 1956 is the starting point for almost all analyses of economic growth. This dynamical model is still used in today's macroeconomic theory despite its relative simplicity. Solow's purpose was to develop a model to describe the dynamics of the growth process and the long-term evolution of the economy \cite{Solow1956contribution,50thSolow}. The Solow model describes the dynamics of the gross domestic product based on the variables labor, capital and technology. Thus, the classical Solow model can estimate the evolution of the physical capital over time through a differential equation. On the other hand, Solow model has been used in other areas such religion, ecommerce, internal brain drain, evaluation of the social benefit of water-saving agriculture and many others \cite{ReligionSolow,ecommerceSolow,SolowBrainDrainItaly,SolowWaterSaving}.

The size and the rate of growth of the economy is of considerable interest for governments and the public in general. The growth of the economy is related to many factors. One important factor for each country is the commercial relations with its neighbors. The smuggling phenomenon is a universal issue, which has been turned into a great problem in many countries. It is inappropriate since it has negative impacts on the economy, weakens the economical strategies and sometimes shows itself as a silent battle against the economy \cite{SmugglingBattle}. Porous borders and weak government capacity are important issues that are faced by several countries \cite{IllicitAmericas}. Furthermore, a large and growing underground economy may have serious economic implications on the countries economic growth. For instance, it can erode the tax base and cause important measurement errors in official economic statistics \cite{Canada2000tobacco}. Because of its hidden nature, the underground economy is very difficult to measure.

Yet from a much broader historical perspective, illicit cross-border flows of various sorts have been a defining feature of U.S. commercial relations with its neighbors from the very start, suggesting that there is much more continuity with the past than conventional accounts recognize. Porous borders and weak government capacity have long defined the region, and attempts to secure borders and tighten controls have often had the perverse and unintended consequence of creating a more formidable smuggling challenge. At the same time, efforts to regulate illicit border crossings have expanded the reach of central government authority and stimulated the development of border enforcement infrastructure and capacity. Bringing this history back in to contemporary policy debates can offer fresh perspectives and lessons and provide an antidote to the often shrill and hyperbolic public discourse today about "out of control" borders.\cite{SmugglingUganda}

%Venezuela, officially the Bolivarian Republic of Venezuela, is organized as a federal republic. Venezuela is an emerging/developing nation (International
%Monetary Fund). Much of the Venezuelan economy is dominated by and dependent upon the petroleum industry, which accounts for approximately 95\% of export earnings, approximately 40\% of the country�s federal budget revenues, and approximately 12\% of GDP (The World Fact Book). The country's currency, the Venezuelan bolivar fuerte (VEF) has lost a significant amount of its value against the US dollar on the official and black market in the last 30 years\cite{EntrepeneurVnzla}. Moreover, in the last 12 months the smuggling problem in Venezuela has been increasing dramatically due mainly to a non-official exchange rate that is now more than ten times the official rate. Venezuela has been one of the largest oil producing countries in the world. Thus, for many years, Venezuela's economic policies, growth, and other related activities have been largely influenced by the petroleum industry.

On the other hand, petroleum product subsidies have increased in recent years. Many countries did not fully pass through the sharp increases in international petroleum product prices that occurred in 2007 and early 2008, resulting in a marked increase in subsidies. After declining along with oil prices during the second half of 2008, subsidies have again started to rise, renewing concerns about the fiscal costs. These concerns have been reinforced by the need in many countries to formulate an exit strategy from the recent crisis-related accumulation of public debt \cite{PetroleumSubsidies}. A subsidy by definition is any measure that keeps prices consumers pay for good or product below market levels for consumers or for producers above market. Subsidy means benefit given by the government to individuals or businesses whether in form of cash, tax reduction or by reducing the cost of goods and services. The purpose of subsidy is to help individuals and businesses purchase/acquire essential goods and services that they may not be able to afford, under normal circumstances. Subsidies take different forms and some have a direct impact on price. These include grants, tax reductions and exemptions or price controls \cite{SubsidyFuel}. Fuel subsidy is applied in oil producing countries, such as: Venezuela, Iran, Saudi Arabia, Egypt, Burma, Malaysia, Kuwait, China, Taiwan, South Korea, Trinidad and Tobago, and Brunei \cite{SubsidyRemovalNigeria}.

Generally, there are two main government subsidies: fuel subsidy where consumers pay a fraction of the price that consumers are supposed to pay and price controls regarding several important products including food. Subsidy, in economic sense, exists when consumers of a given commodity are assisted by the government to pay less than the prevailing market price of same. \cite{FuelSubsidyNigeria}

Based on the previous aspects some countries such as Venezuela are facing an extraordinary smuggling of subsidized petroleum products and other regulated priced products through the borders. This economic behavior is usual in countries with lower prices in comparison with their border countries. The countries with higher prices are often substantial beneficiaries through cross-border smuggling. When relevant, governments should also highlight that subsidies promote smuggling, shortages and black markets \cite{UnequalSubsidiesDeveloping}. As has been mentioned, the supply and demand for smuggled goods depend on interregional price disparities in the presence of a trade ban \cite{IllicitAgri}. The smuggling through the borders of Venezuela has been increased in the last few years. People from Venezuela, Colombia and Brazil often participate in the informal circulation of legal and illegal goods across the border. This economic behavior is also presented in other places. Due to the discrepancy between local understanding of legality and national laws, many potentially newsworthy illegal exchanges are not addressed in the media \cite{IguazuSmuggling}. There are transborder roads between Colombia and Venezuela. However, the main one connects the Venezuelan city of San Antonio with  C\'ucuta, the capital of the Colombian province  Norte de Santander. The Sim\'on Bol\'ivar international bridge can be crossed by foot, car or bicycle. Moreover, no documents are required and there are few inspections through the bridge. On the Colombian side, small-scale vendors of gasoline, known as \emph{pimpineros}, peddle the fuel along the highway, without safety precautions, from small plastic containers. A gallon (3.8 litres) of gasoline costs the equivalent of 4.92 dollars in Colombia, compared to just eight cents in Venezuela, because of the heavy subsidies that make gasoline in this country among the cheapest in the world \cite{IPSNews2012}. The price differential allows 1,000 percent profits to be made from smuggling gasoline. Moreover, nowadays it is not only gasoline, but also food and medicines, subsidized in Venezuela, that are trafficked.

Fuel smuggling to Colombia costs Venezuela US 1.4 billion per year, with 30,000 barrels smuggled daily, according to a report coming out of a bilateral meeting between the Colombian and Venezuelan governments in August, 2013 \cite{venezuelanalysis}. In addition, fuel smuggling to other countries such Brazil, and the islands of Curazao, Aruba, and Trinidad and Tobago are also made. Recently, international studies points to the existence of a correlation between globalization and trans-border economic crimes \cite{TransBorderNigeria}. Moreover, some authors have argued that the human use of the single international political economy, which globalization signifies its transition, has receded to the logic of thinking globally and acting locally \cite{TransBorderNigeria}.

The purpose of this paper is to shed some new light on the debate regarding smuggling through the borders of the countries. In this paper we study the effects of the smuggling of goods across the borders of countries. We rely on an extension of the economic Solow model to describe how the smuggling affects the economic growth of the countries. The model is based on a parabolic partial differential equation that describes the dynamics of physical capital and the boundary conditions are of Neumann type in order to model the physical capital diffusion through the borders of the country. However, there exist many other options that has been proposed to deal with this type of economic issues and is still an open economic debate \cite{NatureEconomyAgent}.

\section{Spatial Solow model}

The Solow model describes the dynamics of the gross domestic product based on the variables labor, capital and technology. Thus, the classical Solow model can estimate the  time evolution of the physical capital through a differential equation. In regard to physical capital there is not complete agreement about what this term refers and how to measure it \cite{PhysicalCapitalDatabase}. In fact some authors have introduced the concept of social capital \cite{SocialCapital}. However, in general for economist, physical capital refers to a factor of production (or input into the process of production), such as machinery, buildings, or computers \cite{samuelson1995economics}. In economic theory, physical capital is one of the three primary factors of production, also known as inputs production function. Another capitals considered in economy are human capital which is the result of investment in the human agent and financial capital. Human capital is the stock of competencies, knowledge, social and personality attributes, including creativity, embodied in the ability to perform labor so as to produce economic value \cite{samuelson1995economics}.

As has been stated in many articles, the mathematical expression for the total production or output in terms of the input variables labor $L(t)$, physical capital $K(t)$ and technology $A(t)$ is given by,
\begin{equation}
Y(t)= A(t)\,F(K(t),L(t)),
\end{equation}
where $F$ denotes the production function. On the other hand, the stock of capital depreciates over time at a constant rate $\delta$ and only a fraction of the output $(cY(t)$ with $0 < c < 1)$ is consumed and $sY(t)$ is the savings (s=1-c). Thus, the Solow model describes the change of the physical capital as,
\begin{equation}
\frac{dK(t)}{dt}= sY(t)-\delta K(t).
\end{equation}
In addition, using the production function one gets,
\begin{equation}
\frac{dK(t)}{dt}= sA(t)\,F(K(t),L(t))-\delta K(t).
\end{equation}
Now if we introduce a change of variable $k(t)=\frac{K(t)}{L(t)}$ in order to compute the per capita physical capital one obtains,
\begin{equation}
f(k(t)= F(K(t)/L(t),1).
\end{equation}
Thus, one gets that,
\begin{equation}\label{model0}
\frac{dk(t)}{dt}= sA(t)\,f(k(t))- (\delta+n(t))\, k(t),
\end{equation}
where $n(t)$ denotes the labor growth rate. Notice that the labour-augmenting technology or technology progress $A(t)$ affects positively the variation of physical capital and, depreciation and labor growth rate affects negatively.

There are several variations that may be included for the based Solow model (\ref{model0}). For instance if we consider a net flow of physical capital to a given location or space interval we need to extend the Solow model in order to account for the flow \cite{CamachoSpatialSolowModel,SpatialSolowTech,CapitalAccumulationSpace}. Thus, $k(x,t)$ denotes the physical capital stock held by a representative household located at x at the time t, in a bounded domain $\Omega \in \mathbf{R}^{n}$. Moreover, here we consider that the production function $f(k(t))$ is the same whatever is the location x. In addition, the initial physical capital distribution, $k(x,0)$, is assumed to be known and $C^{0}$. Thus, one gets the following model for the dynamics of the physical capital,

\begin{align}\label{model}
&\frac{\partial k(x,t)}{\partial t}-\Delta k(x,t)=sA(x,t)\,f(k(t))-\delta k(x,t),\,\,\Omega\times[0,T]\notag\\
&\nabla k(x,t)=h(k(x,t)),\,\,\partial\Omega\times[0,T]\notag\\
&k(x,0)=k_0(x),\,\,x\in\Omega.
\end{align}
This model has been called a spatial Solow model \cite{CamachoSpatialSolowModel,SpatialSolowTech}. Notice that here we introduce a Neumann type boundary conditions in order to deal with the flow of physical capital through the borders of the country in an illegally way. Moreover, the boundary conditions are related in $\partial\Omega$ with the initial physical capital distribution in order to have a well posedness model. In addition, the model (\ref{model}) is still open since the production function $f(k(t))$, the technological progress function $A(x,t)$ and the initial physical capital distribution $k_0(x)$ are unknown.

Several economic growth models are based on production functions. These functions play a crucial role in the economy and different types of function have been used. It specifies the maximum output for all possible combinations of input factors and therefore determines the way the economic model evolves in time \cite{50thSolow,Solow1956contribution,SpatialSolowTech}. The Cobb-Douglas production function is especially notable for being the first time an aggregate or economy-wide production function had been developed, estimated, and then presented to the profession for analysis; it marked a landmark change in how economists approached macroeconomics \cite{ProductionCobbDouglasEastern}. Furthermore, the function was not developed on the basis of any knowledge of engineering, technology, or management of the production process. Nowadays, the Cobb-Douglas production function is by far the most used to represent the technological relationship between the amounts of two or more inputs, particularly physical capital and labor, and the amount of output that can be produced by those inputs. The Cobb-Douglas production function is non-negative, increasing and concave, and, satisfies the well-known Inada conditions \cite{InadaConditions,ProductionCobbDouglasEastern}. However, other productions functions may be used in conjunction with economic models such the Solow model. For instance Leontief production or nonconcave functions are accepted in the applied literature. Here we rely on a particular production function introduced in \cite{SpatialSolowTech}. This function is called an S-shaped production function (nonconcave) allowing the Solow model to have a richer dynamics which includes the existence of a poverty trap. As it has been mentioned in \cite{SpatialSolowTech}, the complexity of the Solow model using the nonconcave function suggest to do the analysis by means of numerical techniques. Thus, for our numerical simulations we consider the following nonconcave production function
\begin{equation}\label{pf}
f(k)=\frac{\alpha k^{p}}{1+\beta k^{q}},
\end{equation}
where the parameters $\alpha$, $\beta$, $p$ and $q$ are usually determined by adjusting the production function by least squares to economic data. In regard to the technological progress function $A(x,t)$ many mathematical expressions have been used. However, a constant value has been assumed in several works. For instance in \cite{CapitalAccumulationSpace} authors used a function such the economy has a technological center, that may coincide with the geographical center. On the other hand, in \cite{SpatialSolowTech} it has been assumed that the technological progress can be modeled by a diffusion process coupled to the physical capital diffusion. Thus, these authors assume that $A(x,t)$ is an unknown to be determined. In our numerical simulations of the next section we assume for the function $A(x,t)$ different mathematical expressions including constant in space and time, and varying linearly or exponentially with time as has been suggested in the literature.

There are lots of evidence regarding smuggling activities in many countries however, there are no reliable statistics available which can be quite misleading \cite{SmugglingBattle}. Numerical simulations are performed in next section in order to study different economic scenarios.  As it has been mentioned the parameter values are difficult to estimate for each country and general numerical values are assumed without loss of generality. In regard to the initial physical capital distribution $k_0(x)$, we use in the numerical simulations four different types. The first one is a constant but is only consistent when the function $h(x,t)$ involved in the Neumann type boundary conditions is equal to zero. The other three types are consistent with $h(k,x)\ne 0$. The first one has a Gaussian form with higher values of physical capital in the middle of the physical space. The last two functions are given by parts, where one is composed by linear forms and the last one is more complex since the function $h(k,x)$ depends explicitly on the physical capital available on the border. More details regarding these mathematical expressions are given in the next section.

\section{Numerical simulations}

We perform several numerical simulations of the spatial Solow model for different scenarios in order to study the effect of physical capital diffusion through the borders of the countries. The proposed model is based on a parabolic partial differential equation and the boundary conditions are of Neumann type. The numerical results are obtained using an explicit finite difference scheme and small time steps in order to fulfill the Courant Friedrich Lewy condition for the numerical stability of the solution. However, implicit methods without stability problems to numerically solve the spatial Solow model can be found in \cite{CamachoSpatialSolowModel,SpatialSolowTech,CapitalAccumulationSpace}.

We study four important factors of the spatial Solow economic model such the depreciation rate, technological progress, initial conditions and boundary conditions. For all these factors we use different numerical values or type of functions in order to observe the effect of each one on the physical capital dynamics. We simulate several unit times in order to observe transient and steady states. It is important to remark that in this first study we do not rely on real world data since our first main goal here is to develop the methodology and analyze the effect of physical capital diffusion through the borders of the countries. Moreover, real world numerical values are very difficult to estimate and for some variables such the level of technology is not observable and estimated values need to be used \cite{SolowLucasOECDpanel}. On the other hand, estimation for the production function also would be necessary and a huge amount of economic panel data is required which is out of the scope of this study.

Here we show the behavior of the physical capital under different scenarios regarding the boundary conditions which are related to the borders of the countries. For each scenario numerical simulations are performed in order to understand the effect of each boundary condition on the dynamics of the physical capital. For our numerical results we consider the S-shaped nonconcave production function (\ref{pf}) with $\alpha=0.0005$, $\beta=0.0005$, $p=4$ and $q=4$. Finally, it is important to mention that a spatial dimension of $L=100$ is used for all the numerical simulations without loss of generality and a numerical value for the saving rate $s=1$ is taking for simplicity on the analysis of the dynamics. It is well-known that the saving rate $s$ in the long term has no effect on the growth rate of the economy \cite{SpatialSolowTech}.

\subsection{No physical capital flux through the borders}

In this first case we consider the classical spatial Solow model without flux on the borders of the countries. In addition, we consider here that the initial condition for the physical capital is given by a uniform distributed function $k(x,0)=100$ and that no technological progress is present, i.e. $A(x,t)=1$. The boundary conditions in explicit form for the physical capital model (\ref{model}) are given by,

\begin{equation}\label{bc1}
\frac{\partial k(0,t)}{\partial x}=0 \hspace{1cm} \text{and} \hspace{1cm}  \frac{\partial k(L,t)}{\partial x}=0.
\end{equation}

\begin{figure}[h]
\centering
\begin{tabular}{cc}
\includegraphics[width=0.45\textwidth]{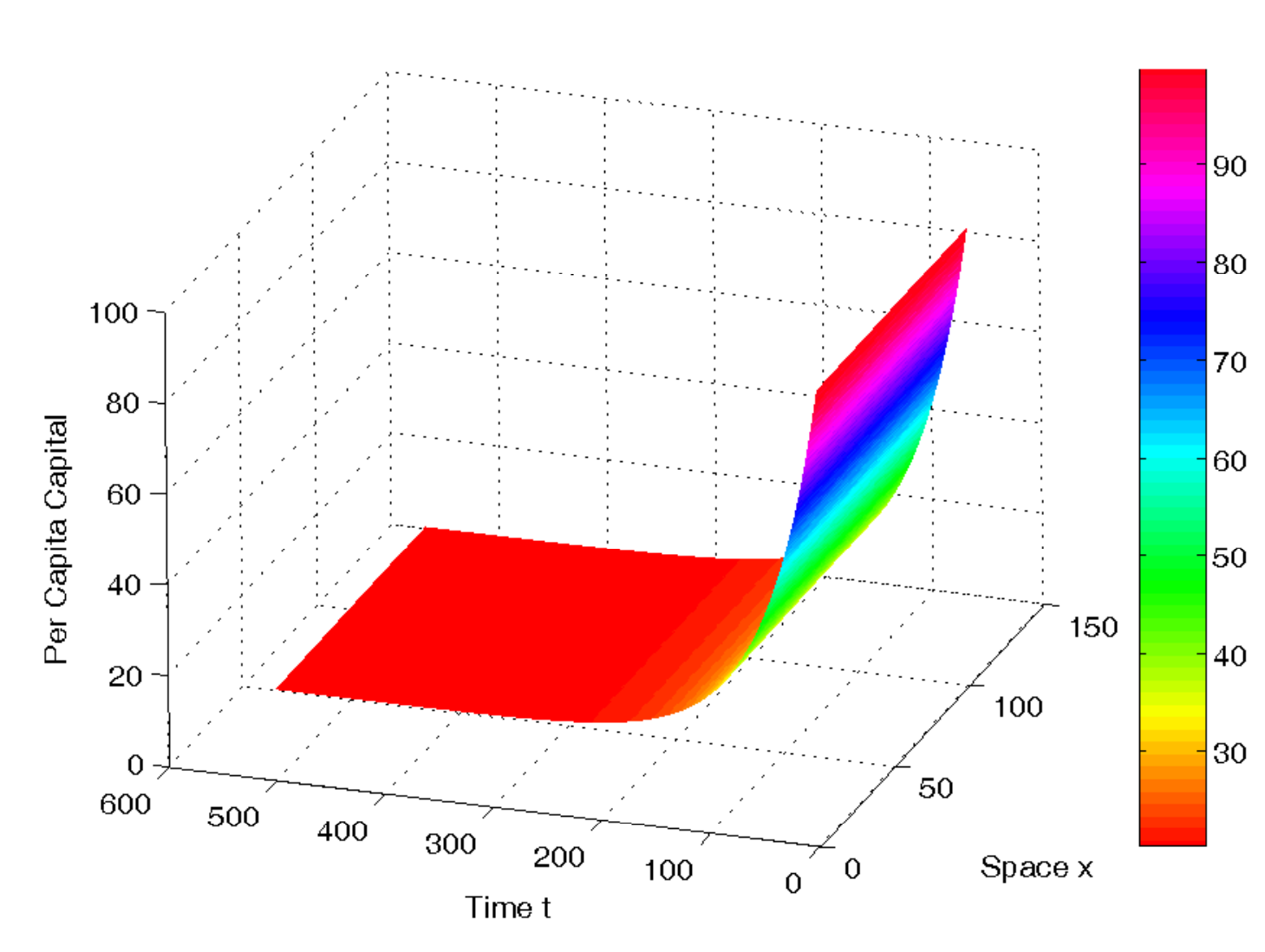}
&\includegraphics[width=0.45\textwidth]{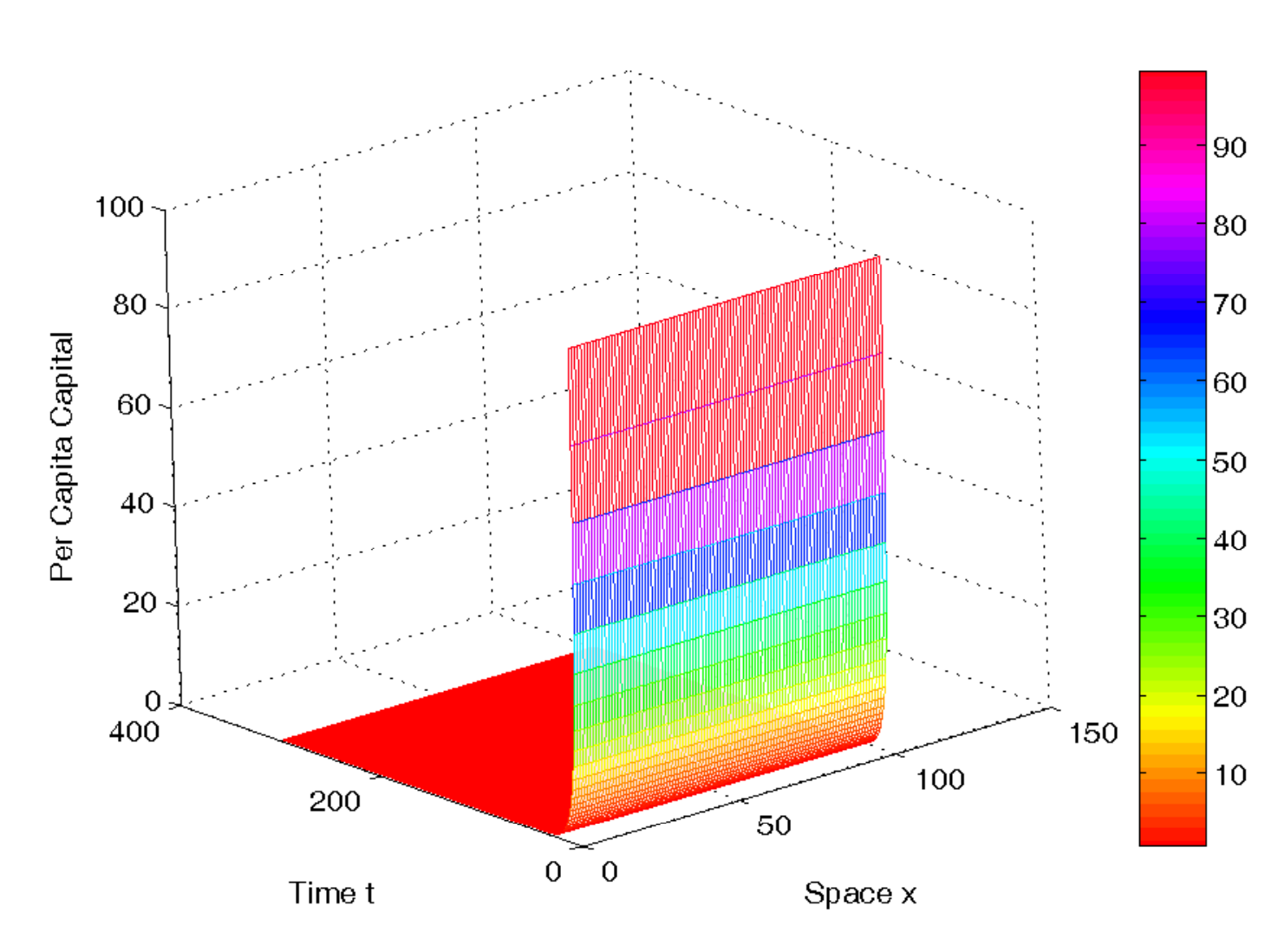}\\
\end{tabular}
\caption{Numerical solutions of the spatial Solow model without physical capital flux through the borders of the country. Initial condition for the physical capita is given by a uniform distributed function $k(x,0)=100$, no technological progress is included here, i.e. $A(x,t)=1$. The increase of the depreciation rate $\delta$ implies a faster decrease of the physical capital as it can be seen on the left ($\delta=0.05$) and right ($\delta=0.5$) hand-sides. }
\label{fig1}
\end{figure}

In these first numerical simulations we would like to point out the effect that the depreciation rate $\delta$ has on the physical capital dynamics. It can be seen in Fig. \ref{fig1} that the physical capital decreases to zero for depreciation rates of $\delta=0.05$ and $\delta=0.5$. However, for the latter case the rate of decrease is much higher due to the depreciation. It is important to remark that one way to avoid the physical capital decrease is to increase the technological progress $A(x,t)$ of the country.

\subsection{No physical capital flux through the borders with technological progress}

\begin{figure}[h]
\centering
\begin{tabular}{cc}
\includegraphics[width=0.45\textwidth]{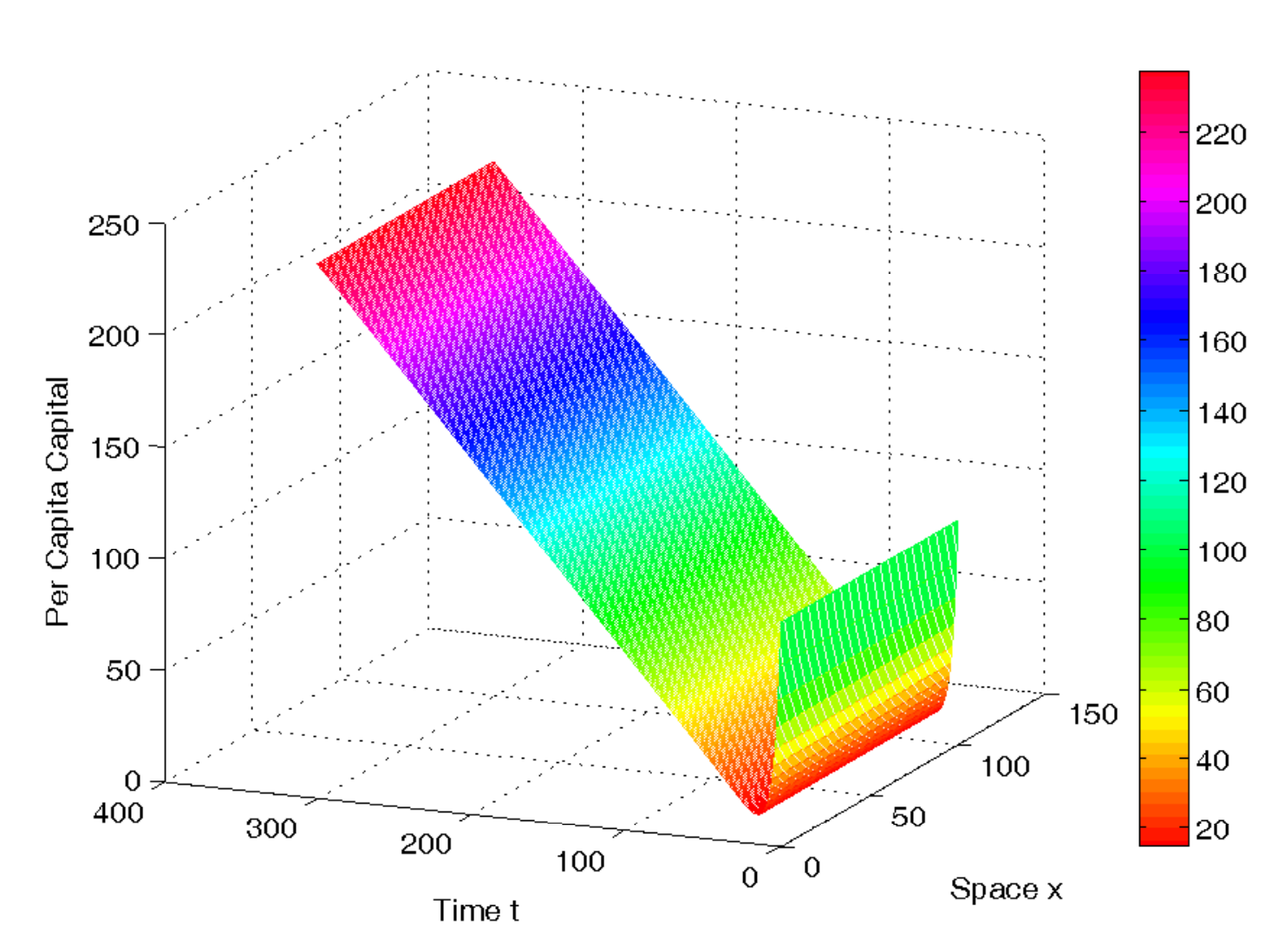}
&\includegraphics[width=0.45\textwidth]{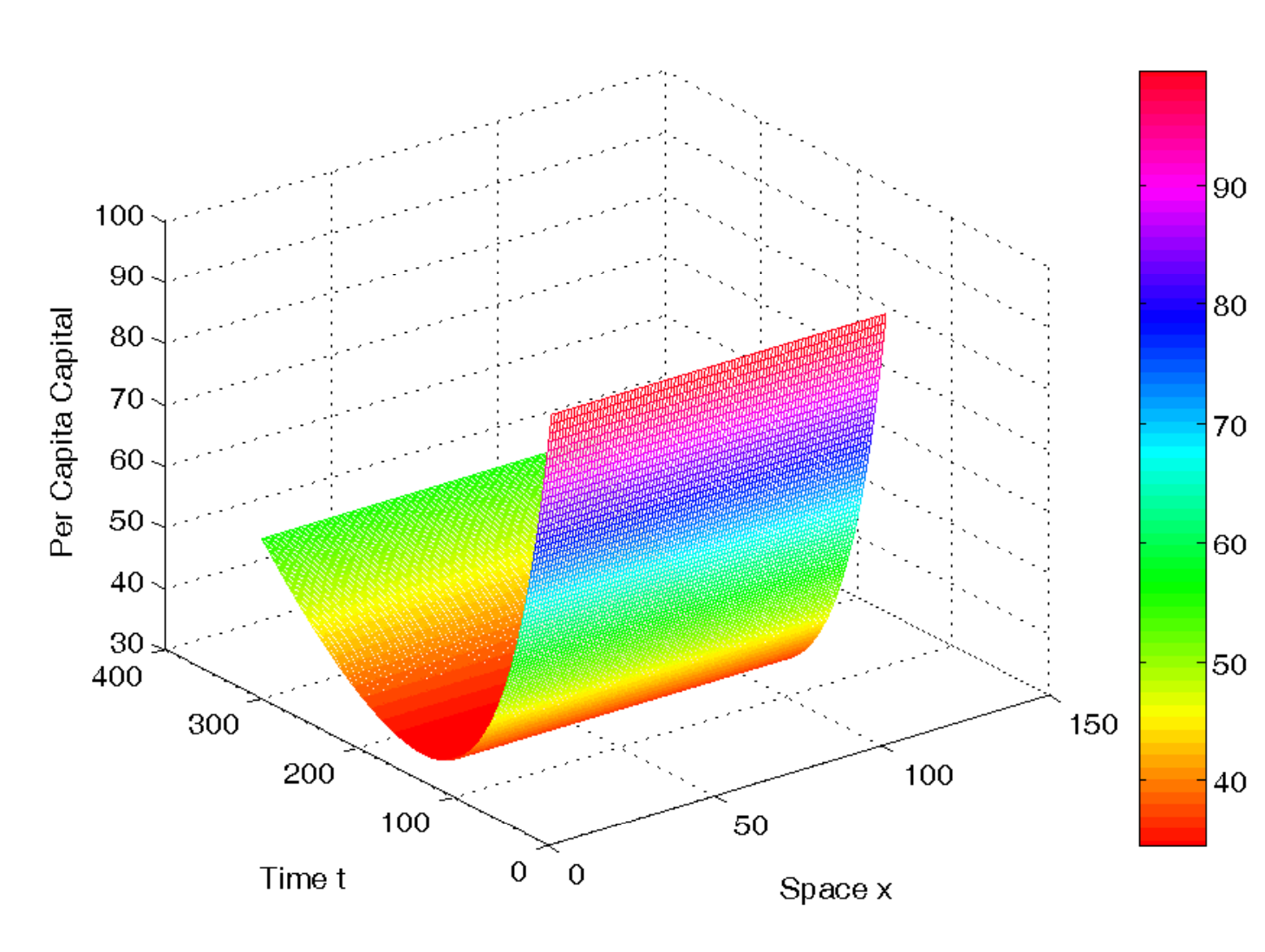}\\
\end{tabular}
\caption{Numerical solutions of the spatial Solow model without physical capital flux through the borders of the country. Initial condition for the physical capita is given by a uniform distributed function $k(x,0)=100$, no technological progress is included here, i.e. $A(x,t)=e^{0.01t}$. The increase of the depreciation rate $\delta$ implies a faster decrease of the physical capital as it can be seen the left ($\delta=0.05$) and right ($\delta=0.5$) hand-sides. }
\label{fig2}
\end{figure}

In this second case we again consider the classical spatial Solow model without flux on the borders of the countries. However, in this case we consider that the technological progress is increasing every year (time). In regard to initial condition we set an uniform distributed function $k(x,0)=100$. For the technological progress we consider two scenarios; the first one with a linear increase $A(x,t)=t$ and the second one with an exponential increase of the technological progress $A(x,t)=e^{0.01t}$. In these scenarios we would like to point out the effect that the technological progress $A(x,t)$ has on the physical capital dynamics. It can be seen in Fig. \ref{fig2} that in both scenarios the physical capital decreases initially and then after some time the physical capital starts to increase due to the technological progress. Notice that in the previous cases, where no technological progress was included, the physical capital drops to zero after some time. Thus, it is important to remark that the technological progress $A(x,t)$ of the country can help for the country's growth, since this variable is related to physical capital \cite{50thSolow,Solow1956contribution,SpatialSolowTech,TechProgressSpatialChina2007,ControlSpatialPollution}. Furthermore, it has been shown that spatial diffusion mechanisms of technology progress allows the develop of other regions and its depends on the spatial distance \cite{SpatialSolowTech,TechProgressSpatialChina2007}. Technological progress may be obtained by investment in human resource, specialization and industrial structure.

\subsection{Physical capital flux through the borders.}

\begin{figure}[h]
\centering
\begin{tabular}{cc}
\includegraphics[width=0.45\textwidth]{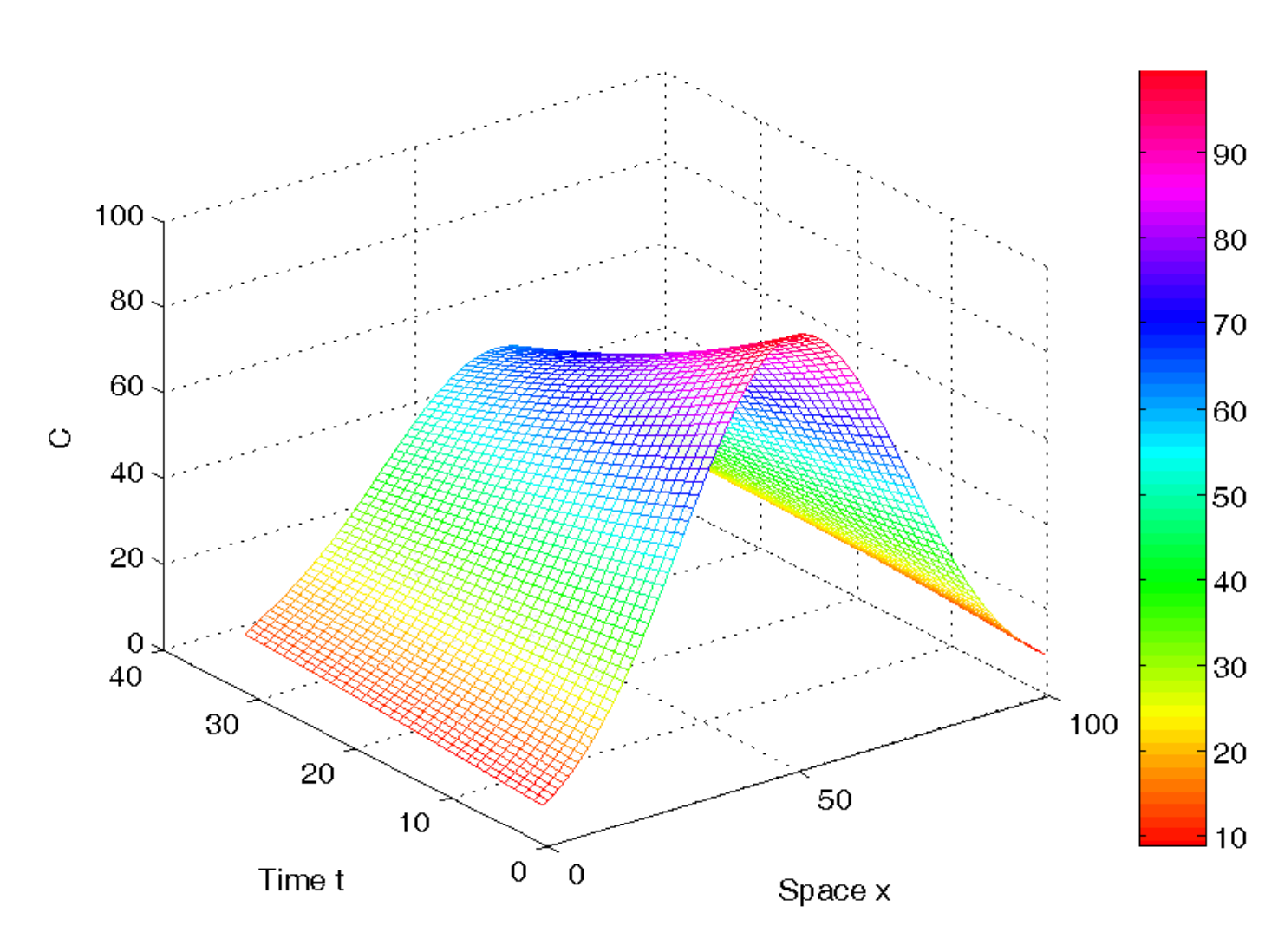}
&\includegraphics[width=0.45\textwidth]{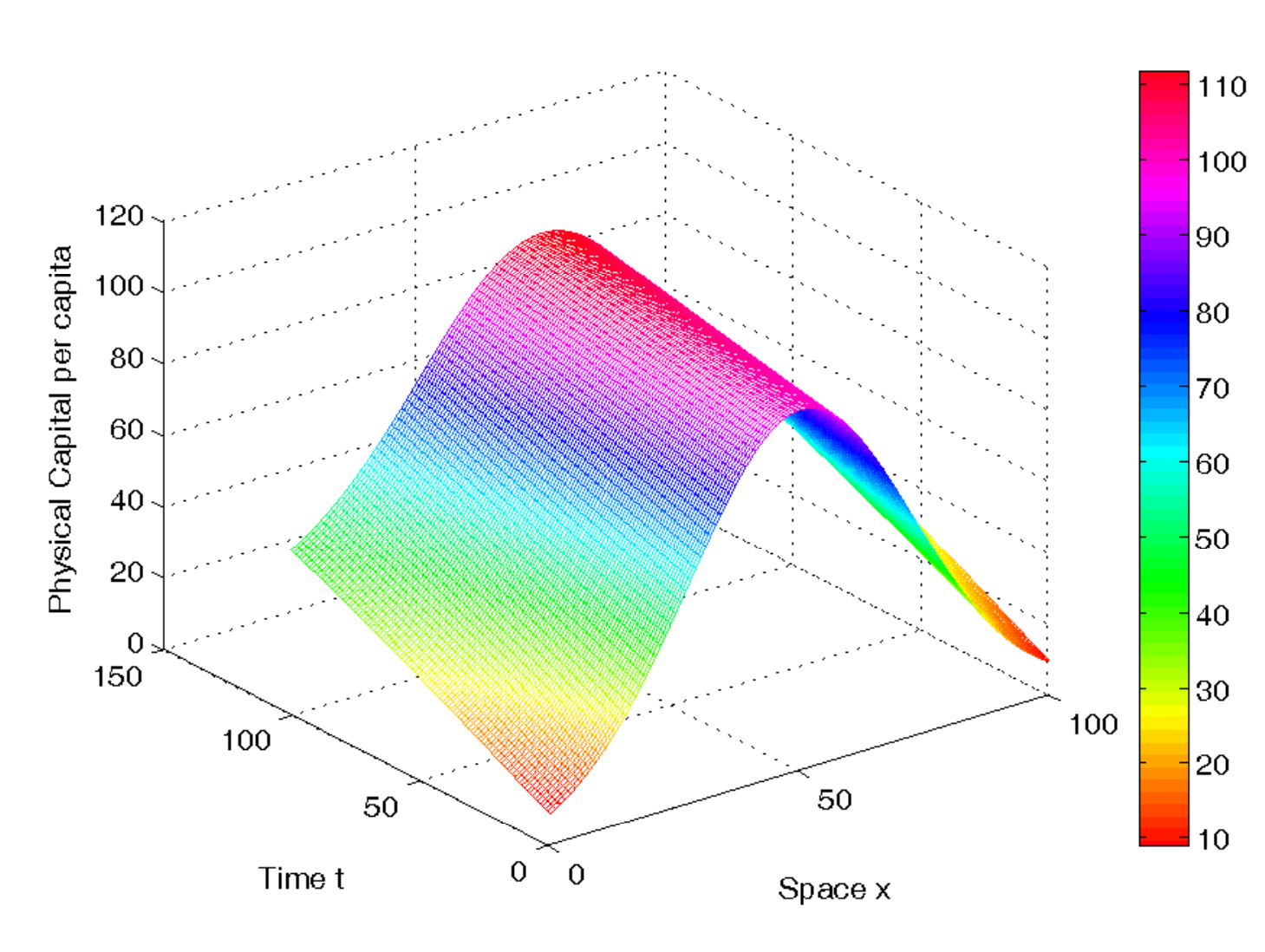}\\
\end{tabular}
\caption{Numerical solutions of the spatial Solow model with physical capital flux through the borders of the country. Initial condition for the physical capital is given by a Gaussian distributed function $k(x,0)=100\,e^{-(x-50)^{2}/1000}$, $A(x,t)=1$ and $\delta=0.05,0.0005$ respectively. }
\label{fig3}
\end{figure}

In this case we would like to study the effect that has the physical capital flux through the borders using the spatial Solow model. In this scenario we consider that the technological progress is constant in order to focus in the effect that has the capital flux through the borders. We focus in three types of boundary conditions in order to simulate the flow of physical capital and such they are consistent with $h(k(x,t))\ne 0$. The first one has a Gaussian form with higher values of physical capital in the middle of the physical space in order to simulate a country where usually the regions near the borders are more exposed to the consequences of the smuggling. For the first case we consider the following boundary conditions,

\begin{equation}\label{bc2}
\frac{\partial k(0,t)}{\partial x}=D_{0} \hspace{1cm} \text{and} \hspace{1cm}  \frac{\partial k(L,t)}{\partial x}=-D_{0},
\end{equation}
where the parameter $D_{0}= 10e^{-(50^2)/1000}$ measures the amount or intensity of physical capital flow through the borders of the country. In regard to initial condition we set a Gaussian distributed function $k(x,0)=100\,e^{-(x-50)^{2}/d}$, where $d$ is a parameter that measures the differences between the central region and the border regions, where a large value of the parameter $d$ means small differences for the region's physical capital. Thus, initial physical capital decreases as we depart from the center. It is important to notice that initial condition must be different than a constant uniform distribution like in the previous cases in order to have a consistency with the physical capital flux through the borders. In other words, the flux induces a nonuniform distribution of the physical capital. Here we consider also two scenarios with different deprecation rates $\delta=0.05$ and $\delta=0.0005$. It can be seen in Fig. \ref{fig3} that when a depreciation rate is high the physical capital will decrease steadily to zero. However, for a lower value of the depreciation rate the physical capital will increase despite the flux of it through the borders. However, due to the physical capital flux now a lower depreciation rate is required in order to avoid a decrease for the capital. Thus, it is important to remark that the flux of physical capital through the borders of the country can undermine the country's growth.

\begin{figure}[h]
\centering
\begin{tabular}{cc}
\includegraphics[width=0.45\textwidth]{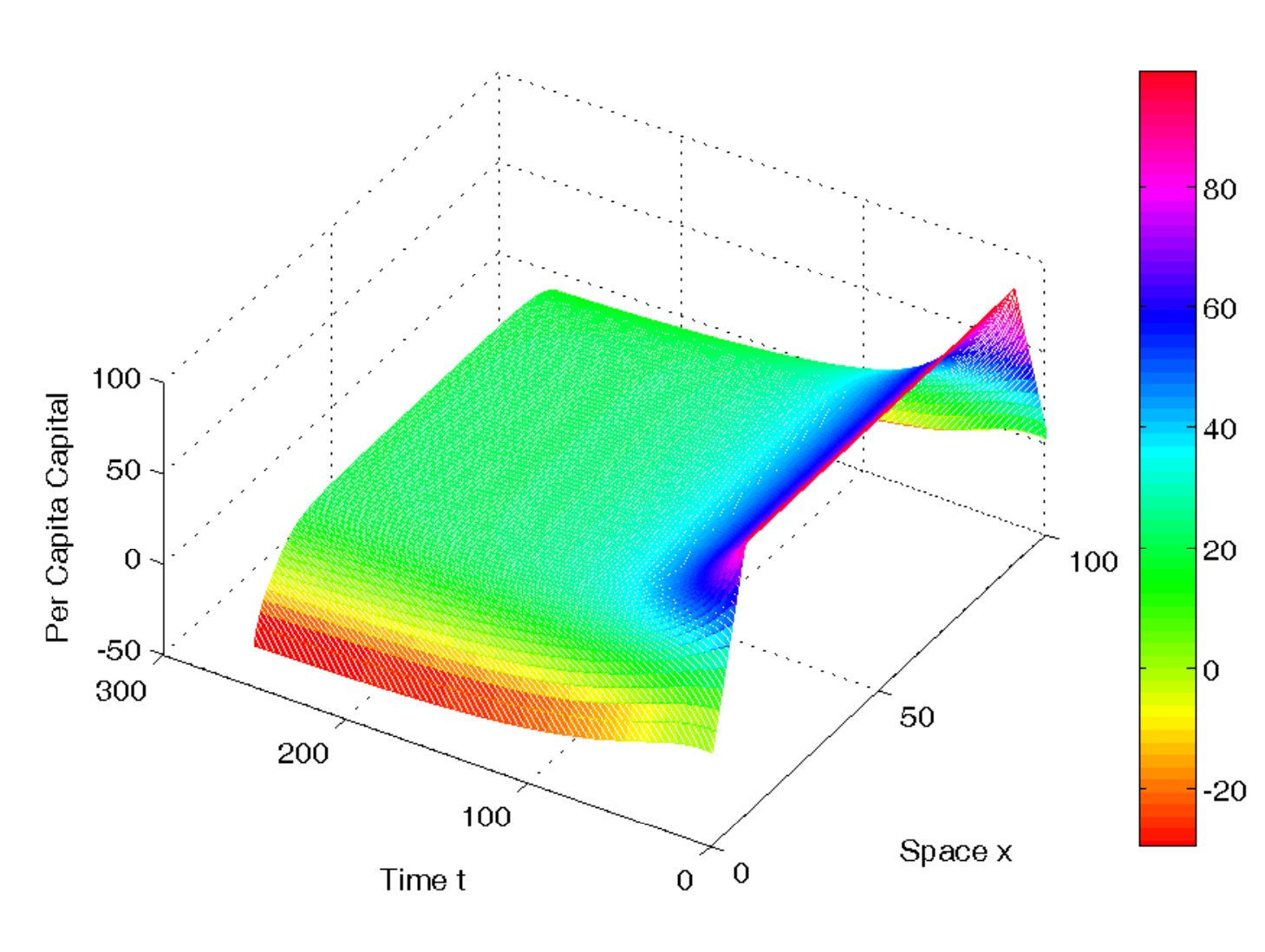}
&\includegraphics[width=0.45\textwidth]{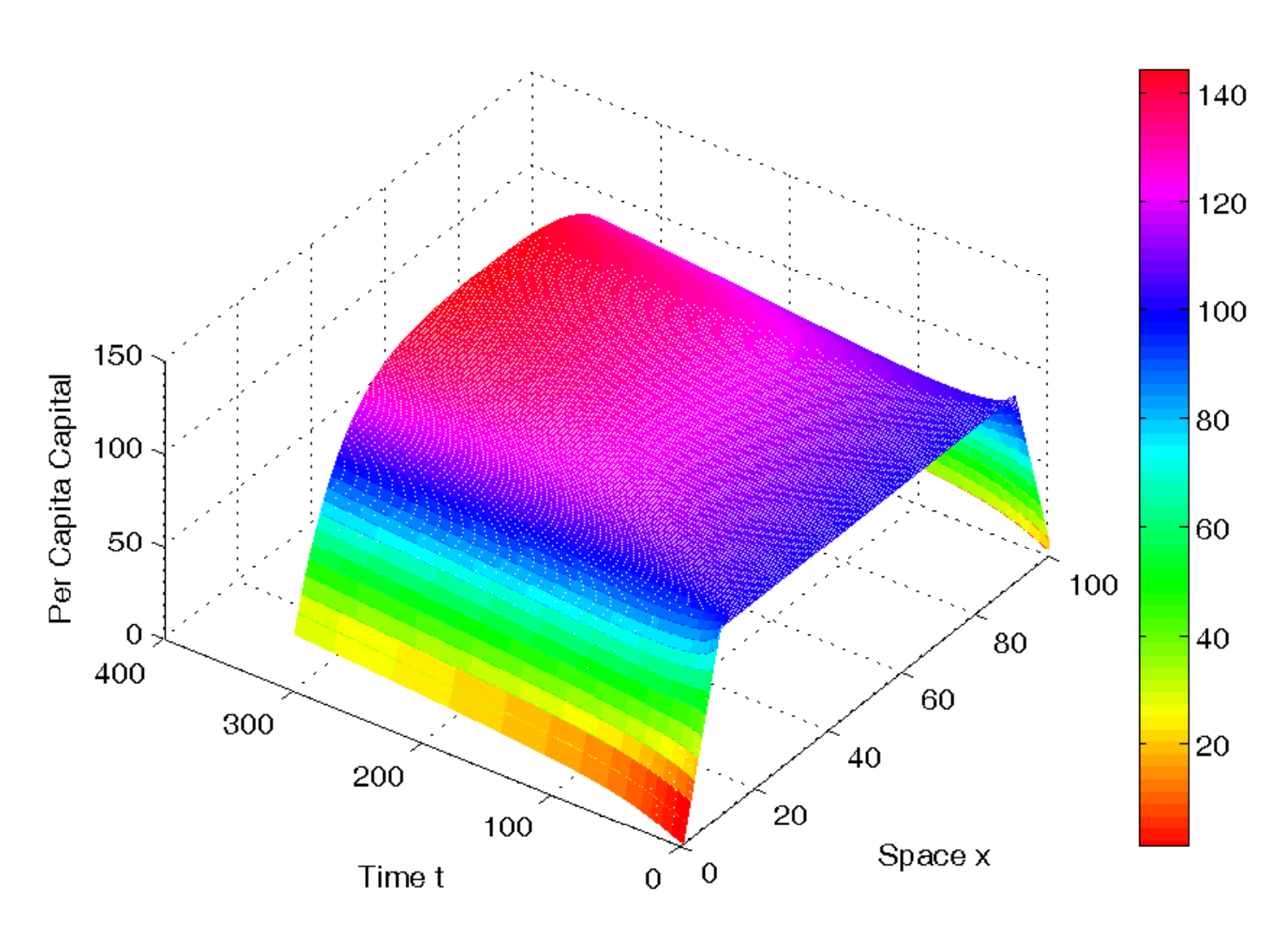}\\
\end{tabular}
\caption{Numerical solutions of the spatial Solow model with physical capital flux through the borders of the country. Initial condition for the physical capita is given by a function given by parts, $A(x,t)=1$ and $\delta=0.05,0.005$ respectively. }
\label{fig4}
\end{figure}

In the second scenario with physical capital flux we consider that the initial condition is given by the following function given by parts,
\[ k(x,0) = \left\{
\begin{array}{c l}
mx & \quad \mbox{if $0\leq x \leq 10 $}\\
C_{max} & \quad \mbox{if $10\le x \le 90$ }\\
-m(x-90)+C_{max} & \quad \mbox{if $90\leq x \leq 100.$}\\
\end{array} \right. \]
where we assume $C_{max}=1000$ and $m=100$. It is important to notice that this physical capital initial condition varies with the position on the regions near the borders and in the central regions is assumed constant in order to simulate as close as possible a real scenario. Moreover, this initial condition is consistent with the boundary conditions that allow the flux of physical capital through the borders. For this case we consider the following boundary conditions,
\begin{equation}\label{bc3}
\frac{\partial k(0,t)}{\partial x}=m \hspace{1cm} \text{and} \hspace{1cm}  \frac{\partial k(L,t)}{\partial x}=-m,
\end{equation}
where the parameter $m$ measures the amount or intensity of physical capital flow through the borders of the country. In order to study the effect of this type of initial condition we also include the effect of the deprecation rate with two numerical values $\delta=0.05$ and $\delta=0.005$. It can be seen in Fig. \ref{fig4} that for the largest depreciation rate the physical capital will decrease steadily to zero. On the other hand, for the smallest value of the depreciation rate the physical capital increases despite the flux of it through the borders. Notice that the physical capital is larger in the central regions than in the border regions due to the flux on the borders.
%In this way, as has been mentioned previously the flux of physical capital through the borders is an economical issue that may affect Venezuela's economy and its growth.

In our last scenario we introduce a more complex boundary condition in order to avoid possible negatives values for the physical capital at the long term and to analyze the effects that has the physical capital flux, depreciation and the initial condition on the economy(physical capital). In this scenario we consider that the initial condition is the following function given by parts,
\[ k(x,0) = \left\{
\begin{array}{c l}
e^{Dx} & \quad \mbox{if $0\leq x \leq 10 $}\\
e^{10D} & \quad \mbox{if $10\le x \le 90$ }\\
e^{D(100-x)} & \quad \mbox{if $90\leq x \leq 100.$}\\
\end{array} \right. \]
where we set the parameter $D=1$. This initial condition is consistent with the boundary conditions that allow the flux of physical capital through the borders. For this case we consider the following boundary conditions,
\begin{equation}\label{bc4}
\frac{\partial k(0,t)}{\partial x}=Dk(0,t) \hspace{1cm} \text{and} \hspace{1cm}  \frac{\partial k(L,t)}{\partial x}=-Dk(0,t),
\end{equation}
where the parameter $D$ measures the fraction of physical capital that flow through the borders of the country.

\begin{figure}[h]
\centering
\begin{tabular}{cc}
\includegraphics[width=0.45\textwidth]{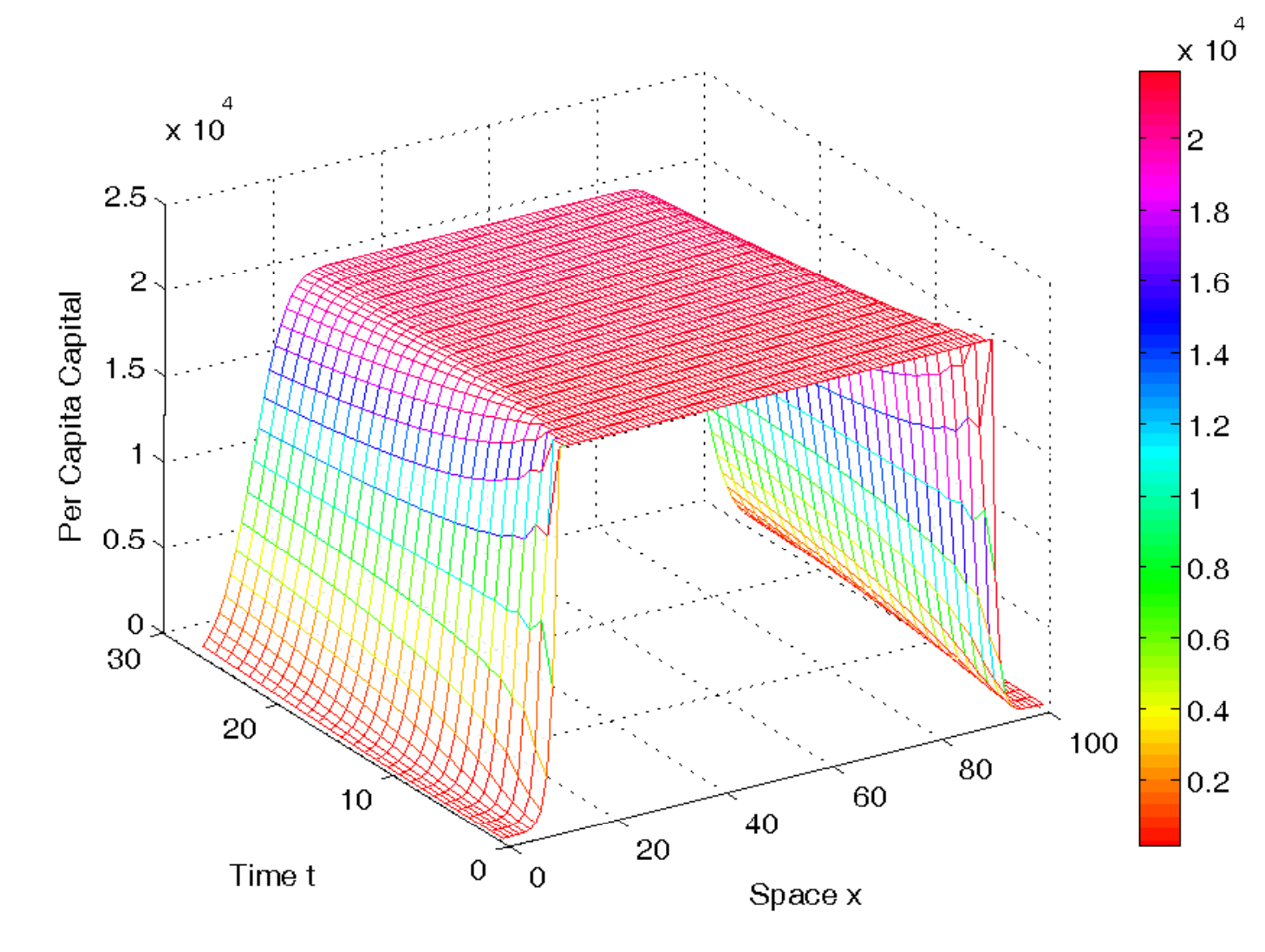}
&\includegraphics[width=0.45\textwidth]{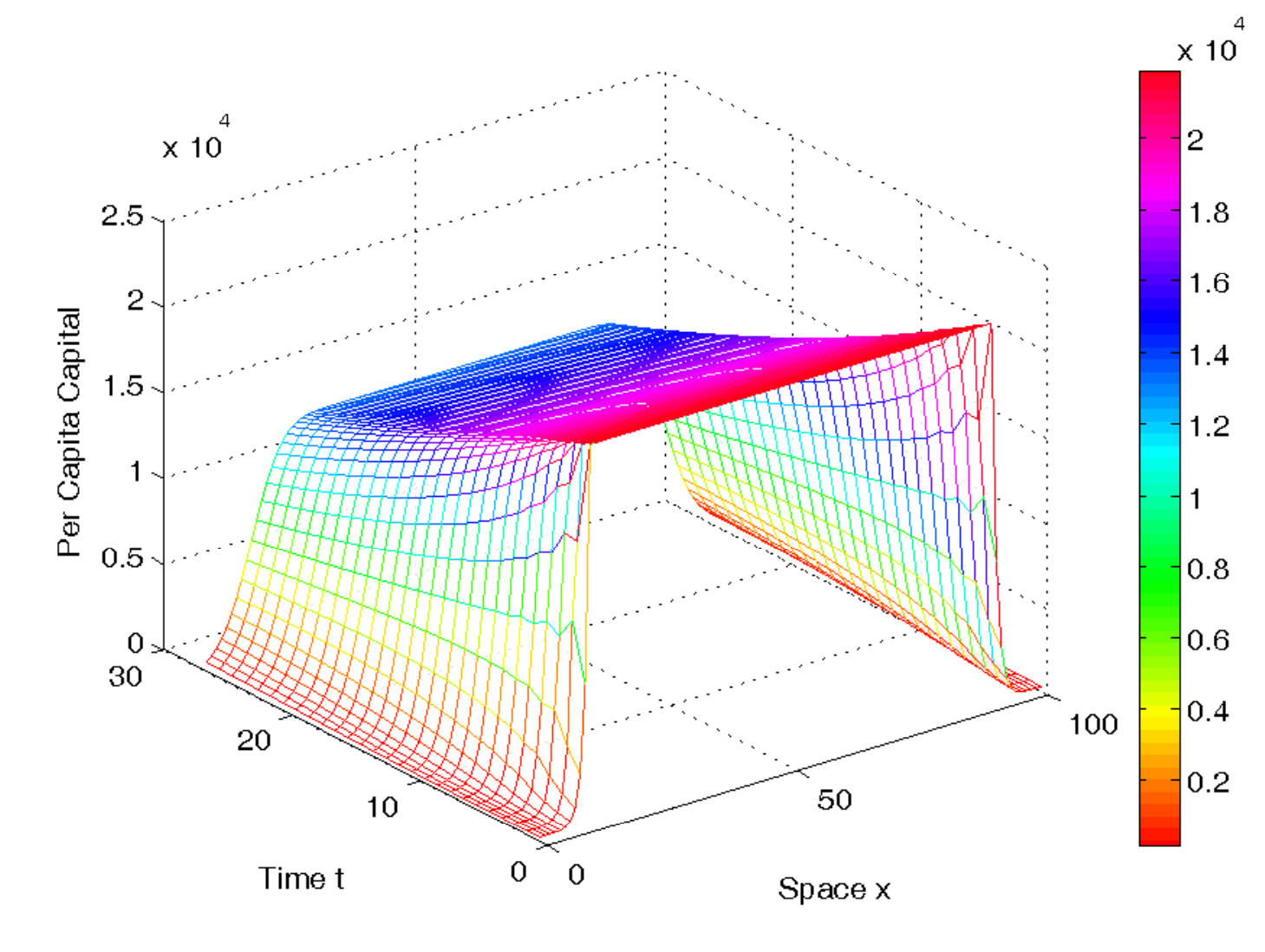}\\
\end{tabular}
\caption{Numerical solutions of the spatial Solow model with physical capital flux through the borders of the country. Initial condition for the physical capita is given by a function given by parts, $A(x,t)=1$ and $\delta=0.05,0.005$ respectively. }
\label{fig5}
\end{figure}

In order to study the effect of the deprecation rate with take two numerical values $\delta=0.05$ and $\delta=0.005$. It can be seen in Fig. \ref{fig5} that for the largest depreciation rate the physical capital decreases in with a lower rate in comparison with the case for the smallest value. Notice again that the physical capital is larger in the central regions than in the border regions due to the flux on the borders. In this way, as has been mentioned previously the flux of physical capital through the borders is an economical issue that can affects the economic growth of the countries and its growth.

\section{Conclusions}

In this study we proposed specific boundary conditions of Neumann type for the spatial Solow model to explain and understand the effect of physical capital diffusion through the borders of the countries. The physical capital has been considered an important variable for the economic growth of a country. In the last decade fuel, medicine, machinery and food smuggling through the borders of the country has been a problematic issue for the Venezuelan economy. Moreover, in the last 12 months the smuggling problem in Venezuela has been increasing dramatically due mainly to a non-official exchange rate that is now more than ten times the official rate.

Here we have used an extension of the economic Solow model to describe how the smuggling of fuel, machinery, goods and others elements affects the economic growth of the countries. In this study we have relied on a production function that is non-concave instead of the classical Cobb-Douglas production function. The proposed model was based on a parabolic partial differential equation that describes the dynamics of physical capital and the boundary conditions  were of Neumann type in order to model the  physical capital diffusion through the borders of a country. The numerical results were obtained using an explicit finite difference scheme that shows how the diffusion of physical capital to the border countries is a paramount issue for the economic growth of the countries.

We can summarize the results using four important factors such the depreciation rate, technological progress, initial conditions and boundary conditions. We found that when there is not out flux of physical capital the country would need less effort regarding the variables depreciation rate, technological progress to achieve an increase of capital. On the other hand, when flux of capital is considered thorough the borders of the country, it is more demanding to increase the physical capital since a lower depreciation rate and an increasing function for the technological progress may be necessary for the economic growth of the country.

Finally, we found that initial condition for the distribution of the physical capital in conjunction with the boundary condition have some effect on the dynamics of the physical capital on the long term. This fact differs with the scenario when there is not flux since at the long term the physical flux is distributed uniformly. Thus, it is important to remark that the flux of physical capital through the borders of the country can undermine the country's growth. Furthermore, numerical simulations showed that the physical capital is larger in the  central regions than in the border regions due to the flux on the borders. In this way, we can conclude that the flux of physical capital through the borders is an economical issue that may affect Venezuela's economy and its growth.

Further research in this economic issue should lead to improve the understanding of the economic problem that many countries face. Moreover our results can help government economic institutions and population to deal and propose economic policies to tackle the flux of physical capital through the borders of the country such as smuggling of fuel, machinery and food. The model also gives likely future outcomes as well as the implications of alternative economic policies. A natural continuation of this research is the extension to the Ramsey model and the introduction of different production functions.\\
%
%Moreover, the results show that the dynamics of the physical capital when boundary conditions of Neumann type are different than zero differ from the classical economic behavior observed in the classical spatial Solow model without physical capital flux through the borders of countries.

\appendix

\section{Appendix}

In order to compute the numerical simulations of problem (\ref{model}), we define the domain discretization as $\Omega_0=[0,L]\times[0,T]$. We discretize the domain $\Omega_0$ in the following form: we chose positive integers $N,T_0$ such that the spatial step size is given by $\Delta x=L/N$, while the time step size is $\Delta t=T/T_0$, and we put $x_j=j\Delta x,$ $t_n=n\Delta t,$\, $0\leq j\leq N,$\, $0\leq n\leq T_0.$ We seek approximations of
the solution at these mesh points, these approximate values will be denoted by $ k_j^n \approx k(x_j,t_n)$. By replacing the first derivative and the second order space derivative by the classical difference quotient it follows that
\begin{align}\label{derivatives}
&\frac{\partial k(x_j,t_n)}{\partial t}=\frac{k(x_j,t_{n+1})-k(x_j,t_n)}{\Delta t}+O(\Delta t),\notag\\
&\frac{\partial^2k(x_j,t_n)}{\partial x^2}=\frac{k(x_{j+1},t_n)-2k(x_j,t_n)+k(x_{j-1},t_n)}{\Delta x^2}+O(\Delta x^2).
 \end{align}
For the sake of simplicity, we define the difference operator
\begin{align}\label{equation99}
    \delta_j^n:=\frac{k_{j+1}^n-2k_j^n+k_{j-1}^n}{\Delta x^2},\qquad 1\leq j\leq N.
\end{align}

Using (\ref{derivatives}) and (\ref{equation99}), it obtains an explicit standard numerical scheme for the problem (\ref{model}), such that for $0\leq n\leq T_0,\ 0\leq j\leq N,$
\begin{align}\label{equation100}
&k_j^{n+1}=(1-\Delta t \delta)k^n_j+\Delta t  \delta_j^n+\Delta t s A^n_j+f(k_j^n),\,\, \Omega \times[0,T],\notag\\
&k^0_j=k_0(x_j),\,\, x_j\in\Omega.
\end{align}

Now, to calculate the values of the operator $\delta_j^n$ in (\ref{equation99}) for $j=0$, and $j=N$ requires us to invoke the fictitious values $k_{-1}^{n}$ and $k_{N+1}^{n}$. These approximations are determined using the boundaries conditions, such that for this case are:
\begin{align}\label{equation101}
\frac{\partial k(0,t)}{\partial x}=d_0h_1(k(0,t)),\qquad\frac{\partial k(L,t)}{\partial x}=d_Lh_2(k(L,t)).
\end{align}
Next, approximating the equations (\ref{equation101}) with central differences scheme, one gets for $k_{-1}^{n}$ and $k_{N+1}^{n}$, the following expressions
\begin{align}\label{equation102}
k_{-1}^{n}=k_{1}^{n}+2d_0\Delta xh_1(k_0^n),\,\quad
k_{N+1}^{n}=k_{N-1}^{n}-2d_L\Delta xh_2(k_N^n).
\end{align}
Thus, we calculate in the boundary points using
{\small \begin{align}
&k_0^{n+1}=(1-\Delta t \delta)k^n_0+2\Delta t  \frac{k_1^n-k_0^n+d_0\Delta xh_1(k_0^n)}{\Delta x^2}+\Delta t s A^n_0+f(k_0^n),\\
&k_N^{n+1}=(1-\Delta t \delta)k^n_N+2\Delta t  \frac{k_{N-1}^n-k_N^n+d_L\Delta xh_2(k_N^n)}{\Delta x^2}+\Delta t s A^n_N+f(k_N^n), \, 0\leq n\leq T_0\notag.\label{equation133}
\end{align}}
Thus, we obtain an explicit numerical scheme which is well-known to be numerically stable and convergent whenever $\frac{\Delta t}{\Delta x^2}\le \frac{1}{2}$. The numerical errors are proportional to the time step and the square of the space step.

%%\bibliographystyle{spmpsci}
%\bibliographystyle{spmpsci}
%\bibliography{BiblioVnzlaSolow}

\end{document}